# The Abzooba Smart Health Informatics Platform (SHIP) ™ – From Patient Experiences to Big Data to Insights

**Naveen Ashish, PhD, Abzooba Inc., USA**

**Antarip Biswas, MS, Abzooba Inc., and Indian Institute of Technology Kharagpur, India**

**Sumit Das, MS, Abzooba Inc., India**

**Saurav Nag, BS, Abzooba Inc., India**

**Rajiv Pratap, BTech, Abzooba Inc., USA**

**Abstract**

*This paper describes a technology to connect patients to information in the experiences of other patients by using the power of structured big data. The approach, implemented in the Abzooba Smart Health Informatics Platform (SHIP), is to distill concepts of facts and expressions from conversations and discussions in health social media forums, and use those distilled concepts in connecting patients to experiences and insights that are highly relevant to them in particular. We envision our work, in progress, to provide new and effective tools to exploit the richness of content in social media in health for outcomes research.*

## Introduction

In this paper we describe a platform that we have developed for automatically structuring discussions in health social media forums, with the objective of using the structured information to provide better retrieval and analytic insights to end users. With the increasing use of the Internet as a resource for health information, particularly in social media in health, we are witnessing a very high proliferation of user (patient) discussions on health issues. According to the Pew Internet Project report[1] 78% of Internet users (or 59% of US adults) look up health information online, and about one in five of health information seekers use *social media*. The resources range from purely patient driven discussion forums to question-answer forums with patient as well as qualified medical practitioner participation.

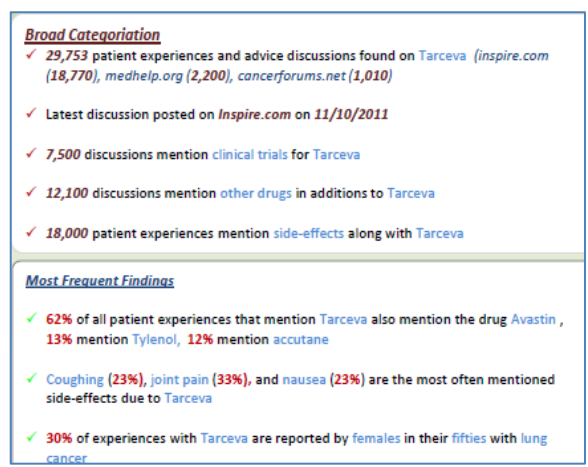

*Figure 1- Insights from Patient Experiences*

Our goal is to provide valuable and effective means to connect patients to the experiences of *other patients like them* using the power of big data distilled from the original information. Figure 1illustrates a subsection (for brevity) of information provided to a patient in response to a search for "Tarceva"(the chemotherapy drug). The 'Broad Categorization' panel provides information such as which forums are most actively discussing Tarceva. The 'Frequent Findings' panel highlights insights such as coughing, joint pain and nausea being the most frequently mentioned *side-effects* in the discussions related to Tarceva. The patient can further (not shown) click on the links and go to such experiences collected from various forums and read them. The analytic capabilities above also open up new possibilities of insights for clinical outcomes research. Indeed, the patient initiated observational study on lithium carbonate use in ALS patients [2] refuting a 2008 published study on the subject, demonstrates the value that information in social media on patient experiences can bring to outcomes research.

This paper primarily describes the Abzooba Smart Health Informatics Platform or SHIP – which has a distillation pipeline for automatically distilling structured *facts and expressions* from the experiences of patients. The structured facts and expressions go into an *experiences meta-base* (Fig 2), on top of which powerful retrieval capabilities can be provided.

## Methods

### Distillation Pipeline

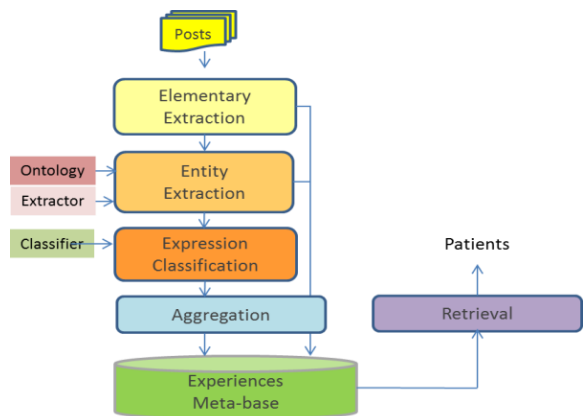

*Figure 2 - Distillation Pipeline*

At the heart of the SHIP is a distillation pipeline illustrated in Figure 2, driven by technologies such as machine learning, automated natural language processing and ontologies [2,3]. Posts of patient experiences (for consistency we will refer to an entire discussion as a 'Thread' and individual postings in that thread as 'Post') go through various extraction and classification in the pipeline and finally through an aggregation step (explained shortly),by the end of which we have distilled about 50 facts and expressions from such post. These facts and expressions go into a structured big data database or what we call our **experiences meta-base**. The steps of the distillation pipeline along with the aggregation module to create a structured meta-base of facts and expressions from the text in the posts are described:

### Elementary Extraction

The very first step is to perform some basic text processing on each thread and post. This involves providing a unique ID to each post and thread, parsing using the HTML, and extracting HTML based information such as number of replies, when last updated etc.

### Entity Extraction

The next step is to extract medically significant entities relevant to health such as drugs (includes chemotherapy drugs and regimens, pain medications, and other drugs), side-effects, treatments and procedures, adverse events, hospitals or locations, etc. We have used the *XAR* system which in turn leverages open-source systems (such as "GATE" and "UIMA") [2]. We further built upon XAR by integrating in ontologies from theUMLS (Unified Medical Language System) [3] and other lexicons and ontologies for entities such as side effects and pain medications.

### Expression Distillation

The third step in the concept distillation pipeline is to classify each post to one of {Y,N} (yes or no) depending on whether the post contains a particular expression of interest or not. Specifically we have designed and implemented classifiers for five expressions of interest: (i) **Personal experience** i.e., whether the post has an expression of someone having personally experienced something of interest (such as having experienced a particular condition, treatment procedure, drug etc.,). (ii) **Advice** i.e., whether there is any advice in the post on something of interest. (iii) **Information** i.e., whether there is any informational reference (Websites, books or other resources) mentioned in the post. (iv) **Support** i.e., whether the message contains an expression of encouragement and/or support (such as *"good luck, my thoughts and prayers are with you!"*). (v) **Outcome** i.e., whether the post is telling about some positive or negative outcome after taking some medicine or following some medical procedure. All the above five classifiers are based on using a "J48" *decision tree* algorithm [4]. The features, provided for classifier induction, vary depending on the type of classification and are largely based on expression phrases and terms in the text. Table 1 shows some representative examples of the various facts, entities and expressions that are distilled for each post.

### Aggregation

Finally the elementary facts, extracted entities and classified expressions extracted at the post level are *aggregated* at the thread level. Aggregation is the process of synthesizing information distilled from posts in a thread to the parent thread. For entities, say age group or cancer stage, a union is taken of all the entities in each post and ascribed to the parent thread.

On the other hand for a classified expression such as advice or personal experience – a presence of such in any single post in a thread ascribes the presence to the parent thread.

Table 1 - Distilled Facts and Expressions

| Elementary Facts |
|---|
| Last updated, Number of responses, Top level category, Post length, URL, Expert authored, User name |
| **Extracted Entities** |
| Age, Gender, Condition, Cancer stage, Location, Cancer condition, Treatment (procedure), General Drug, Chemo drugs, Pain killers, Side-effects, Date diagnosed, Hospital |
| **Classified Expressions** |
| Personal experience, Advice, Information, Support, Outcome |

**Information Retrieval over Experience Meta-base**

As a first step we have developed a *Lucene* [5] based (text) search interface to the information with optimizations added for indexing, dynamic score boosting and local caching.

## Results

### Evaluation

We have applied our system to four major discussion forums – Inspire[1], Medhelp[2] and two other forums, and on data for six major types of cancer (breast, lung, skin, prostate, brain and blood). The base data comprises close to 50,000 discussion threads that in turn comprise of close to 400,000 discussion posts. The resulting experiences meta-base contains over 20 million facts and expressions distilled from this base data. We have further built an initial version of an information retrieval engine over this meta-base using which we can (i) Provide analytic insights (ii) Provide better search through filters.

As a use case, consider patient experiencing severe cough after starting on the chemotherapy drug Tarceva. She looks up the Tarceva Web site[3] as well as a site on a clinical trial related to Tarceva. While both these sites talk about the primary side effects of Tarceva, cough is not mentioned at all (as a possible side effect). A search on "Tarceva" on our site however provides the information on the analytic dashboard that cough is actually the third *most common side effect* mentioned in the discussions on Tarceva ! This indicates to her that her cough may actually be due to taking Tarceva, a fact confirmed by her perusal of some of the experiences (under "Cough"). She could further filter and hone in on discussions that contain some advice on how to deal with cough while on Tarceva.

### Distillation Accuracy

The extraction accuracies, in terms of precision and recall, for each of the elementary facts and extracted entities are in the range of 0.9 i.e., 90%. Table 2 below provides the classified expressions accuracies (first letter of classifier for brevity). By design we have placed a higher emphasis on precision vs recall for challenging cases.

Table 2 – Classifier Accuracies

| Classifier | P | A | I | S | O |
|---|---|---|---|---|---|
| **Precision** | 0.87 | 0.91 | 0.93 | 0.89 | 0.80 |
| **Recall** | 0.82 | 0.62 | 0.91 | 0.90 | 0.58 |

### Related Work

There are several efforts in progress on technologies for automated distillation of information from (natural language) text data in the medical domain [6]. Our work is distinguished from such efforts in that (i) We distill expressions such as for personal experience, advice and outcome in posts, which is not addressed by prior work (ii) We provide selective insights over the aggregated data in an automated fashion.

**Address for correspondence**

Naveen Ashish, 365 San Juan Place Pasadena CA 91107 USA. naveen.ashish@abzooba.com Tel: 408 757 7284.

---
[1] http://www.inspire.com
[2] http://www.medhelp.org
[3] http://www.tarceva.com